\documentclass{aa}

\newcommand{\msun}{\mbox{$M_{\odot}$}}
\newcommand{\lsun}{\mbox{$L_{\odot}$}}
\newcommand{\rsun}{\mbox{$R_{\odot}$}}

\newcommand{\teff}{\mbox{$T_{\rm eff}$}}
\newcommand{\vinf}{\mbox{$v_{\infty}$}}
\newcommand{\vesc}{\mbox{$v_{\rm esc}$}}
\newcommand{\ratio}{\mbox{v$_{\infty}$/v$_{\rm esc}$}}
\newcommand{\mdot}{\mbox{$\dot{M}$}}

\newcommand{\mster}{\mbox{$M_{\star}$}}
\newcommand{\msunyr}{\mbox{$M_{\odot}$y$^{-1}$}}
\newcommand{\kms}{\mbox{km~s$^{-1}$}}

\usepackage{graphics}
\begin{document}

\thesaurus{08(08.05.1; 08.05.2; 08.13.2; 08.18.1; 08.19.3; 08.23.3)}
\title{The radiation driven winds of rotating B[e] supergiants}

\author{Inti Pelupessy\inst{1}
 \and Henny J.G.L.M. Lamers\inst{1,2}
 \and Jorick S. Vink\inst{1}
}

\institute{ 
    {Astronomical Institute, University of Utrecht,
    Princetonplein 5, NL--3584 CC, Utrecht, The Netherlands.
    {\tt f.i.pelupessy@phys.uu.nl; lamers@astro.uu.nl; j.s.vink@astro.uu.nl }}
 \and
    {SRON Laboratory for Space Research, Sorbonnelaan 2,
     NL--3584 CA, Utrecht, The Netherlands}
         }
\offprints{H.J.G.L.M. Lamers}
\mail{Astronomical Intitute, Princetonplein 5, NL-3584~CC, Utrecht,
    The Netherlands}

\date{Received 9 February 2000 / Accepted ???} 

\maketitle

\begin{abstract}

We have formulated the momentum equation for sectorial line driven 
winds from rotating
stars including: (a) the oblateness of the star, (b) gravity
darkening (von Zeipel effect), (c) conservation of angular momentum, (d) line driving
specified by the force multiplier parameters
($k$, $\alpha$, $\delta$), (e) finite disk
correction factors for an oblate  star with gravity darkening for both the
continuum and the line driving. The equations are solved numerically.
We calculated the distribution of the mass flux and the wind velocity 
from the pole to the equator for the winds of B[e]-supergiants.
Rotation decreases the terminal
velocity in the equatorial region but hardly affects the wind velocity from
the poles; it 
enhances the mass flux from the poles while the mass flux from the
equator remains nearly the same.
These effects increase with increasing rotation rates.

We also calculated models with a bi-stability jump around 25~000 K, using force
multipliers recently calculated with a Monte Carlo technique.
In this case the mass flux increases drastically from the pole to the
equator and the terminal velocity decreases drastically from pole to
equator. This produces a
density contrast in the wind $\rho({\rm equator})/\rho({\rm pole})$ of about
a factor 10 independent of the rotation rate of the star. 
We suggest that the observed density contrast of a factor 
$ \sim 10^2$ of the disks of B[e] stars
may be reached by taking into account the wind compression due to the
 trajectories of the gas above the critical point
towards the equatorial plane.

\keywords{stars: early type -- stars: emission line, Be
 --stars: B[e] -- stars: mass loss
-- stars: rotation -- supergiants -- stars: winds, outflows}
\end{abstract}


\section{Introduction}
\label{s_intro}

In this paper we study the effects of rotation on the radiation
driven winds of early-type supergiants. We will focus on the explanation for
the occurrence of disks around fast rotating B[e] supergiants.

B[e] supergiants, also designated sgB[e] stars (Lamers et al.~\cite{la98}),
are B type supergiants that exhibit forbidden emission lines in
their optical spectra.
The observations of hybrid spectra, i.e. spectra with {\it broad} UV P Cygni
features and {\it narrow} emission lines and dust emission of B[e]
supergiants in the Magellanic Clouds led to the proposal of a disk wind model
by Zickgraf et al.(\cite{zick85}). This model postulates a dense disk of 
outflowing material in a fast line driven wind to explain the
observed characteristics of the spectra of these stars (see Zickgraf ~\cite{zick92}). 
The stellar wind at the equator is about ten times slower than that at the pole. Also,
the wind at the equator is about a hundred times denser than at the pole. 
Additional evidence for a two-component outflow has been obtained from polarimetric 
measurements, e.g. by Zickgraf \& Schulte-Ladbeck~(\cite{zsl89}).

The precise mechanism behind the formation of these disks is still a
mystery (Cassinelli 1998).
It is however clear that the origin of an axisymmetric wind structure such as
a disk may well be connected to the fast rotation of a star. 
Two theories using rotation in a different way have been considered for the 
formation of these disk winds: 

-- (1) The Wind Compressed Disk (WCD) model of
Bjorkman \& Cassinelli ~(\cite{bc93}), that invokes the kinematics of
the winds from rotating stars. The streamlines of the gas in the wind
from both hemispheres of a rapidly rotating star cross in the
equatorial plane. The concentration of the gas and the shock in the
equatorial plane produce an outflowing equatorial disk, with a thickness on
the order of a few degrees. Owocki \& Cranmer (1994) have argued that
the motion to the equatorial plane may be counteracted by the radiation
force perpendicular from the plane due to lines. 
However, this ``wind-inhibition effect''
may not be effective in winds with a strong density gradient
in the equatorial direction. We return to this in Sect. 7.

-- (2) The rotationally induced bi-stability model (RIB) of Lamers \& 
Pauldrach~ (\cite{lp91})  invokes an increase 
in the mass flux from the equator and a decrease in the equatorial
wind velocity compared to the poles from the bi-stability jump.
This jump in mass flux and in wind velocity is due  
to the temperature difference between the pole and the equator of a fast 
rotating B[e] supergiant with gravity darkening. The jump will occur 
for stars with effective temperatures between 20~000 and 30~000 K.

In reality, both effects, i.e. the wind compression and the 
rotation induced bi-stability, may be operating together 
and amplifying  one another in the line driven 
winds of rapidly rotating early-B stars.
(For a detailed
explanation of both models, see Lamers \& Cassinelli (1999)
(hereafter $ISW$), Chapter 11.)

The theory of radiation driven winds for non-rotating stars
was developed by Castor et al. (\cite{cak})
(hereafter CAK) and predicts the mass loss rate and $\vinf$ for spherical
winds. The influence of rotation on line driven winds was 
investigated by Friend \& Abbott (\cite{fa86}). They found an increase 
in the mass-loss rate at the equator and a corresponding decrease in $\vinf$,
but these authors did not consider the effects of gravity darkening and the
oblate shape of the star. Cranmer \& Owocki ~(\cite{co94}) described the finite disk
correction factor in case of an oblate star.

In this paper we will modify the line driven wind theory,
including these effects of oblateness and gravity darkening as well as the rotational
terms in the equation of motion. This will then be applied to rapidly 
rotating B[e] supergiants to investigate whether these effects can
explain the occurrence of disks.
We will also apply recent bi-stability calculations to study the
effect of the RIB-model to explain the occurrence of outflowing disks 
around rapidly rotating B[e] supergiants.

In Sect. 2 we describe the theoretical background of the radiation driven
wind theory and the bi-stability effect. In Sect. 3 we derive the
equations for the winds from oblate rotating stars with a temperature
gradient between poles and equator due to gravity darkening.
In Sect. 4 we describe the method for  solving the equations and for calculating
the winds from rotating stars. These calculations will be applied to 
B[e] supergiants in Sect.~\ref{s_B[e]}. In Sect.~\ref{s_bs} we will 
investigate the effect of bi-stability on rotating stars and 
Sect.~\ref{s_concl} concludes with a summary of this work and a discussion on
the formation of disks of B[e] supergiants.


\section{Theoretical context}

\label{s_physics}

The computation of the dynamics of line driven stellar winds is a complicated problem in
radiation hydrodynamics. It involves the simultaneous solution of the
equations of motion, the rate equations and the radiative transfer equations 
in order to calculate the radiative acceleration.

For hot luminous stars the line forces are the most important driving forces in the
wind. CAK have shown that the line acceleration can be
parameterized in terms of the optical depth parameter 
$t \sim \rho (dr/dv)$ as $g_{\rm line} \sim k t^{\alpha}$,
where $k$ and  $\alpha$  are parameters that depend on composition and
temperature of the wind. In this expression
$k$ is a measure of the number of lines and $\alpha$ is a measure of the 
distribution of the line strengths with $\alpha=0$  or 1 for 
a pure mix of optically thin or thick lines respectively.

In this paper we adopt the CAK formalism and simplify the
equations of motion by assuming a stationary, radial flow 
of a viscousless fluid.  The possible influence of magnetic fields
will be ignored.
These simplifications are subject to the following restrictions
(see e.g. Abbott 1980):
\begin{itemize}
\item
The CAK theory treats the line acceleration in a simplified way in terms
of the force multipliers, described above. Detailed
calculations of the line driving, including millions of lines,
show that the total line acceleration can be well described by this
simple representation (Pauldrach et al. 1986). 
\item
Both observations and calculations show that line driven stellar winds
are not stationary. Even a wind in a stationary solution will 
develop shocks (see e.g. Lucy~\cite{lu82}). However, the time-averaged structure
 follows the stationary state quite well (see Owocki et al.~\cite{ow88}, and Feldmeier
1999).
Therefore we will restrict ourselves to stationary models.
\item
A sectorial model will be adopted. This means that for every latitude a one dimensional problem
will be solved. Wind compression as in the WCD model will thus be neglected, but the main
effect of wind compression is expected to {\it redistribute} the mass loss and not to change the
{\it total} mass that is lost from the rotating star. 
\item
The absence of viscosity is a good approximation at the high 
density of line driven winds (see CAK). 

\end{itemize}

The representation of the line force by a simple power law
may seem to be a gross simplification of
the underlying physics of myriad line absorption processes, but it can be shown
to hold for a homogeneously distributed mix of optically thick and thin
lines (Abbott~\cite{ab82}, Gayley~\cite{ga95}). Calculations of CAK and more recently in NLTE
by Vink et al.~(\cite{vink99}) of realistic model atmospheres, 
based on calculations of unified atmosphere/wind models with
the ISA-WIND code (de Koter et al. 1997), 
confirm this 
to be generally true to good accuracy for the important part of the wind 
(from about the sonic point up to a few stellar radii).
Therefore, the complex physics of ion populations can be ignored in the investigation
of the various effects of rotation on the stellar wind. 

However, there is one 
notable exception which is called the bi-stability jump. This refers to a
jump in $\vinf$ around the temperature of 21000 K where 
$\vinf/v_{\rm esc}$ climbs from 1.3 (lower temperature) to 2.6 (higher
temperature). This jump was observed by Lamers et al.
(\cite{lsl95}) in B supergiants.
It is linked to a shift in ionization states of Fe in
the lower part of the wind. The line driving in the lower part of the wind is dominated
by iron. Below about $\teff \simeq 25 000$ K, Fe {\sc iv} recombines to Fe {\sc iii} and since
Fe {\sc iii} is a more efficient line driver than Fe {\sc iv}, the wind structure changes
dramatically (Vink et al.~\cite{vink99}).

The bi-stability jump may also occur in the temperature difference between the pole 
and the equator of a fast rotating star with gravity darkening. 
Therefore two sets of force multiplier parameters for the wind will be adopted 
to reflect the sudden change in ionization states (a high-temperature set for the pole 
and a low-temperature set for the equator).

\section{The physics of rotation}
\label{s_form}

Intuitively it is clear that rotation has an effect on the shape of a star and
on the motion of the gas in the wind. An additional effect that the rotation
of a star can have is the darkening of the equatorial regions of a star via
the von Zeipel effect.
These effects will modify the wind of a star considerably and we will 
incorporate them in the line driven wind theory for rotating stars.

\subsection{The shape of a rotating star}

The shape of a uniformly rotating star with all its mass concentrated in the core
is determined by the equipotential surfaces of the potential in a rotating frame 
(Roche model):

\begin{equation}
\label{eq_firot}
\Phi(x,\theta,\phi)=\frac{1}{x} + \frac{1}{2} \omega^2 x^2 \sin^2(\theta)
\end{equation}
where $x=r/R_{\rm eq}$, $\theta$ is the co-latitude ($\theta$ = 0 at the pole) and 
$\omega=v_{\rm rot,eq}/v_{\rm crit}$. $\Phi$ is of course independent of the
 longitude $\phi$. Note that $1+0.5 \omega^2=R_{\rm
eq}/R_{\rm pole}$. So the maximum oblateness is $R_{\rm eq}/R_{\rm pole}=3/2$.
The critical (break-up) velocity is defined in this paper as:
\begin{equation}
v_{\rm crit}^2~=~G M_{\star}^{\rm eff}/R_{\rm eq}
\end{equation}                                       
where $M_{\star}^{\rm eff}$ is the effective mass of the star, 

\begin{equation}
M_{\star}^{\rm eff} = M_{\star} ( 1-\Gamma_e)
\end{equation}
which is the reduced mass due to radiation pressure by electron
scattering (see below).
 
The surface of a rotating star is implicitly given by Eq.~\ref{eq_firot}. Solving
this expression for $x(\theta)$ is equivalent to the solution of
a 3rd degree polynomial, which results in

\begin{equation}
\label{eq_rtheta}
x(\theta)=2 \frac{\sqrt{2+\omega^2}}{\sqrt{3}~ \omega \sin{\theta }}
        \sin{\left\{ \frac{1}{3} \arcsin{ \left( 
          \frac{3 \sqrt{3}~ \omega \sin{\theta } }
        {(2+\omega^2)^{3/2}}\right)}\right\} }
\end{equation}
Because of the difference in radius 

\begin{equation}
\label{eq_rratio}
\frac{R_{\rm eq}}{R_{\rm pole}}=1+0.5 \omega^2
\end{equation}
and the difference in
rotational velocity between pole and equator,
the radial escape velocity at the equatorial
region is smaller than at the pole

\begin{equation}
\label{eq_vescp-e}
v_{\rm esc}^{\rm eq} = \sqrt{\frac{1-0.5 \omega^2}{1+0.5 \omega^2}} ~v_{\rm esc}^{\rm pole}
\end{equation}
We can then roughly estimate the effect of rotation on $\vinf$ as this
value has an almost linear dependence on $\vesc$ (see later in \S \ref{s_ssol}).


\subsection{Von Zeipel gravity darkening}

\begin{figure}
\resizebox{\hsize}{!}{\includegraphics{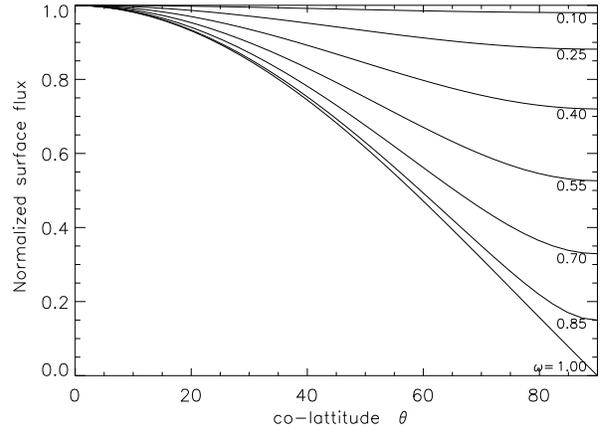}}
\caption{Gravity darkening for different rotation rates $\omega$ as a
function of co-latitude $\theta$}
\label{fg_f}
\end{figure}

The von Zeipel theorem~(\cite{zeipel}) for distorted stars states
that the radiative flux from a point
on the star is proportional to the local effective gravity:
\begin{equation}
\label{eq_zeipel}
F(\theta,\phi)=\sigma_{\rm B} T_{\rm eff}^4(\theta) \propto g_{\rm eff}(\theta)
\end{equation}
with $\sigma_{\rm B}$ the Boltzmann constant. 
As $g_{\rm eff}=-\nabla \Phi$, we can write down the flux of a rotating star as a function
of the co-latitude $\theta$. This ( lengthy) expression for the flux as a function of
$\theta$ is plotted in Fig.~\ref{fg_f} for a few different values of $\omega$.

\subsection{The equation of motion of a line driven wind of a rotating star}

The rotation of the star can induce 
a $\theta$ component of the flow of the matter in the wind. 
In this study a possible $\theta$ component is neglected as was discussed in \S \ref{s_physics},
but the motion in the longitudinal $\phi$ direction must still be taken into account.
 
In the absence of forces that can exert a torque on the wind, the equation
of motion for $v_{\phi}$ is given by the conservation of angular momentum.
Note however that there could be a torque from the line forces themselves, which is 
dependent on the velocity gradient of the gas. 
We assume that there are no external torques, so conservation of angular
momentum gives:

\begin{equation}
v_{\phi}(r)=v_{\rm rot}(R_{\star}) \frac{R_{\star}}{r}
\end{equation}

The equation of motion for the radial direction becomes:

\begin{equation}
\label{eq_bewvgl}
v \frac{dv}{dr}+\frac{G \mster}{r^2}+\frac{1}{\rho}\frac{dp}{dr}-
 v_{\rm rot}^2 \frac{R_{\star}^2}{r^3}- g_{\rm rad}=0
\end{equation}
where $v$ is the radial velocity and $\mster$ is the stellar mass.

The conservation of mass for a non-spherical sectorial wind can be written as

\begin{equation}
\label{eq_massflux}
4 \pi r^2 \rho(\theta,r) v(\theta,r) = F_m(\theta)
\end{equation}
where $F_m(\theta)$ is the ''local mass loss rate,'' i.e. the total mass
 loss rate if the solution for this latitude where valid for a spherical star. 
 The total mass loss rate from the star is 

\begin{equation}
\label{eq_massloss}
\mdot = \int^{\pi/2}_0 F_m(\theta) \sin (\theta) d \theta
\end{equation}

The equation of motion together with the conservation of
mass governs the dynamics of the wind given 
the equation of state, $p=a^2\rho$ ($a$ is the isothermal sound speed), and
the radiative acceleration $g_{\rm rad}$. Using the conservation of mass, the
pressure term can be rewritten as:
\begin{equation}
\label{eq_pres}
\frac{1}{\rho}\frac{dp}{dr}=\frac{1}{\rho}\frac{da^2}{dr}-\frac{2a^2}{r}-\frac{a^2}{v}\frac{dv}{dr}
\end{equation}

As in the CAK-theory the temperature structure is specified a priori 
as a function of $r$. (The results depend only very weakly on the 
chosen temperature structure; see Pauldrach et al.~\cite{ppk86}). 

The radiative acceleration consists of two components. 
The continuum component due to the electron scattering, $g_{e}$, and a second component, $g_{\rm L}$, 
due to line scattering and absorption processes. 
The continuum acceleration is given in terms of $\Gamma_{\rm e}$:
\begin{equation}
\label{eq_cont}
g_{\rm e}= \frac{\sigma_{\rm e} F}{c} =\Gamma_{\rm e} \frac{G \mster}{r^2}
\end{equation}
where $\sigma_e$ is the electron opacity, $L$ is the stellar luminosity and $F$
is the radiation flux. The contributions by other continuum opacities
are  small. Nevertheless, they were included in the calculation of the force
multipliers by Vink et al. (1999) used in our bi-stable wind models
(Sect. 6).

For a homogeneous spherical star, $\Gamma_{\rm e}$ is given by\footnote{Note:
 we define $\Gamma_e$ as the ratio between the continuum force
and gravity and not as the ratio between continuum force and the critical
radiation force for a rotating star}:

\begin{equation} 
\Gamma_{\rm e} = \frac{\sigma_{\rm e} L_*}{4 \pi G \mster}
\label{eq_gam}
\end{equation}
In the more general case of a non-homogeneous, non-spherical star the continuum acceleration
can be defined as a correction to the 'classical' continuum acceleration of Eq.~\ref{eq_gam}:
\begin{equation}
\label{eq_cco}
\Gamma_{\rm e}' =D_{\rm c} \frac{\sigma_{\rm e} L_*}{4 \pi G \mster} 
\end{equation}
where $D_{\rm c}$ is the continuum correction factor which is given by:

\begin{equation}
\label{eq_dc}
D_{\rm c}= \frac{4 \pi r^2}{L_*} \oint_{\rm disk} I(\theta,\phi) d\Omega
\end{equation}
The line acceleration $g_{\rm L}(r,v,v')$
has a more complicated form, since it is a function of distance $r$ as well
as the  velocity $v$ and velocity gradient $dv/dr$.
Combining the equation of motion (Eq.~\ref{eq_bewvgl}) 
with the rewritten pressure and continuum
acceleration terms and multiplying by $r^2$, one finds:

\begin{eqnarray}
\label{eq_bewvglf}
F_{\theta}(r,v,v')  \equiv \left( 1-\frac{a^2}{v^2} \right) r^2 v \frac{dv}{dr} 
      -v_{\rm rot}^2 \frac{R_{\star}^2}{r} \nonumber \\
 +G \mster (1- \Gamma_{\rm e}') - 2a^2 r -r^2 g_{\rm L}=0
\end{eqnarray}

This equation is valid for each co-latitude $\theta$. Following CAK, the 
critical point of this equation is found by imposing the singularity condition 
(where the subscript $c$ indicates values at the critical point):
\begin{equation}
\label{eq_sing}
\left( \frac{\partial F_{\theta}}{\partial v} \right)_c=0
\end{equation}
and the regularity condition:
\begin{equation}
\label{eq_regu}
\left( \frac{\partial F_{\theta}}{\partial r} \right)_c+\left( v' \frac{\partial
F_{\theta}}{\partial v} \right)_c=0
\end{equation}
Note that the critical point is not the sonic point $r_s$ ($v(r_s)=a$) due to
the fact that there is an additional dependence on $v'$ in the line
acceleration $g_{\rm L}$ (Abbott~1980; $ISW$ Chapt 3.3).

\subsection{The radiative line forces}
\label{s_grad}

The line forces are described within the framework of CAK in the Sobolev
approximation. This means that the intristic line absorption profiles are considered to be 
infinitely sharp. Then the line force becomes a function of local properties of the
wind only. The acceleration due to an ensemble of lines is given by
the summation of the contributions of all individual lines with rest
frequencies $\nu_l$. $g_{\rm L}$ is given by (Castor~\cite{ca74}):

\begin{equation}
\label{eq_gmeerlijn}
g_{\rm L}=\sum_{l} \frac{\kappa_{l}}{c} \oint I_{\nu_{l}}
\frac{1-e^{-\tau_{\nu_{l}}}}{\tau_{\nu_{l}}}
\mu d\Omega
\end{equation}
where $\kappa_{l}$ is the absorption coefficient per gram for the $l$th line,
$\mu = \cos{\theta}$,
$I_{\nu_{l}}$ the intensity at the rest frequency $\nu_{l}$ and
$\tau_{\nu_{l}}$ is the
Sobolev optical depth of $l$th line, defined as 
$\tau_{\nu_{l}}=\kappa_{l} \rho\frac{c}{\nu_{l}}(\frac{dr}{dv})$. 
The integration is performed over the complete visible disk of the star.

Following CAK, the line acceleration can be rewritten in terms of the
force multiplier as a function of the optical depth parameter $t$. Where $t$ is defined 
as:
\begin{equation}
t=\sigma_{\rm e}^{\rm ref} v_{\rm th} \rho \frac{dr}{dv}
\label{eq_t}
\end{equation}
with $v_{\rm th}$ being the thermal velocity of the protons
and $\sigma_{\rm e}^{\rm ref}$ is some reference value
for the electron scattering $\sigma_{\rm e}^{\rm ref}= 0.325$
cm$^2$~g$^{-1}$ (see {\it ISW}, Chapt. 8).
The line acceleration in terms of the continuum acceleration is give by:

\begin{equation}
\label{eq_glgem}
g_{\rm L}=g_{\rm e} M(t)
\end{equation}
If the star is assumed to be a point source, $M(t)$ can be approximated
by a simple power law parameterization (CAK, Abbott 1982): 
\begin{equation}
M_{\rm point}(t)~=~k~t^{-\alpha} \left(\frac{n_{\rm e}}{W}\right)^{\delta}
\end{equation}
where the $(n_{\rm e}/W)^{\delta}$ accounts for the effect of the
electron density on the ionization balance in the wind. Here $n_e$ is the electron 
density in units of $ 10^{11}$ cm$^{-3}$ and $W$ is 
the geometric dilution factor.

Friend \& Abbott (1986) and Pauldrach et al. (1986) showed the 
importance of the finite disk correction on the 
line force in case of an extended source. This finite disk 
correction factor $D_{\rm fd}$ is given by (CAK):

\begin{equation}
\label{eq_corr}
D_{\rm fd}=\frac{M(t)}{M_{\rm point}(t)}=\frac{1}{N} \oint I(\theta,\phi) \left(
\frac{(1+\sigma) }{1+ \sigma \mu^2} \right)^\alpha \mu d \Omega
\end{equation}
where

\begin{equation}
\label{eq_sigma}
\sigma=\frac{r}{v} \frac{dv}{dr} -1 ~~~~~ {\rm and} ~~~~~
N= \frac{L_{\star} R_{\star}}{r}
\end{equation}
This expression is completely general: the actual shape of the star and its
intensity distribution enter through the integration domain and the $I(\theta,\phi)$
dependence. However, here we assume the intensity $I$ to be locally isotropic,
i.e. we neglect limb darkening.                
We have tested the effect of limbdarkening. 
The inclusion of limbdarkening results in a small
$\theta$-independent correction on $D_{\rm fd}$.

The line acceleration $g_{\rm L}$ including all the neccesary correction 
factors is now given by:

\begin{equation}
\label{eq_gl}
g_{\rm L}=\left(\frac{\sigma_{\rm e}}{4 \pi}\right)^{1-\alpha} \frac{k L_{\star}}{r^2 c}
(v_{\rm th} F_m)^{-\alpha} D_{\rm fd} \left(r^2 v \frac{dv}{dr}\right)^{\alpha}
\left(\frac{n_{\rm e}}{W}\right)^{\delta}
\end{equation}
Combining Eqs.~\ref{eq_bewvglf} and~\ref{eq_gl} gives the full
equation of motion for the sectorial wind of a rotating star

\begin{eqnarray}
\label{eq_bewvglrot}
0 = \left(1-\frac{a^2}{v^2}\right)r^2 v \frac{dv}{dr}
      + G \mster (1- \Gamma_{\rm e}) - 2a^2 r - v_{\rm rot}^2 \frac{R_{\star}^2}{r} 
        \nonumber \\
  - \left(\frac{\sigma_{\rm e}}{4 \pi}\right)^{1-\alpha} 
      \frac{k L_{\star}}{c}
      (v_{\rm th} F_m)^{-\alpha} D_{\rm fd} 
       \left(\frac{n_{\rm e}}{W}\right)^{\delta}
       \left(r^2 v \frac{dv}{dr}\right)^{\alpha}
\end{eqnarray}



\section{Solutions of the equation of motion}
\label{s_ssol}

For line driven winds of non-rotating stars solutions of the equation of
motion have been retrieved by e.g. CAK and Pauldrach et al. (1986). 
These will be presented to serve as an illustration of the solution
of the equation of motion and as a basis to interpret the more complicated results 
from the full equation of motion (\ref{eq_bewvglrot}).

\subsection{Simplified solutions for non-rotating star}

\subsubsection{The point source approximation}

If the star is considered to be a point soure the integral of the line
acceleration in Eq.~\ref{eq_gmeerlijn}
 collapses to one point. In this case $D_{\rm fd}$ disappears
from  Eq.~\ref{eq_gl} and the line acceleration becomes a simple function
$g_{\rm L}=\frac{C}{r^2} (r^2 v \frac{dv}{dr})^{\alpha}$. 
The solution of Eq.~\ref{eq_bewvglf}
was found by CAK and can easily be explained by neglecting the
gas pressure terms ( $a^2$). In this case Eq. ~\ref{eq_bewvglf} reduces
to

\begin{equation}
\label{eq_bewvglza}
r^2 v \frac{dv}{dr} =G \mster (1-\Gamma_{\rm e})+ C \left(r^2 v
  \frac{dv}{dr}
  \right)^{\alpha}
\end{equation}
where $C$ is a constant containing the mass loss rate.  
The equation is solved by imposing uniqueness of the solution 
(Kudritzki et al. 1989) 
which then fixes the value of the constant and thus $\dot{M}$. 
The solution for the mass loss and velocity law is given by

\begin{eqnarray}
\label{eq_mdotsimple}
\mdot &= & \frac{4 \pi}{\sigma_e v_{th}} \left(\frac{\sigma_e}{4 \pi}\right)
        \left(\frac{1-\alpha}{\alpha}\right)^{\frac{1-\alpha}{\alpha}}
        (k \alpha)^{\frac{1}{\alpha}} \nonumber \\
      &  & \left\{\frac{n_e}{W}\right\}^{\frac{\delta}{\alpha}}
        \left(\frac{L_{\star}}{c}\right)^{\frac{1}{\alpha}}
        \{G \mster (1-\Gamma_{\rm e})\}^{\frac{\alpha-1}{\alpha}}
\end{eqnarray}

\begin{equation}
\label{eq_psopl}
v(r)=\vinf \left(1-\frac{R_{\star}}{r}\right)^{0.5} 
\end{equation}
with 

\begin{equation}
\label{eq_vinfsimple}
\vinf =\sqrt{\frac{\alpha}{1-\alpha}} \vesc =\sqrt{\frac{\alpha}{(1-\alpha)}
\frac{2 G M_{\star} (1-\Gamma)}{R_{\star}}}
\end{equation}
This simplified solution is equal to the full CAK solution, in the
limit of small sound speed, $ a<< \vesc $ (see also $ISW$ Chapt 8).

\subsubsection{Simple finite disk correction}

In case of a homogeneous spherical star the finite disk correction factor 
$D_{\rm fd}$ (Eq.~\ref{eq_corr}) can be calculated analytically (CAK):

\begin{eqnarray}
\label{eq_hd}
D_{\rm fd} &=& \frac{2}{(1-\mu_{\star})} \int_{\mu_{\star}}^{1} \left(
\frac{(1+\sigma) }{1+ \sigma \mu^2} \right)^\alpha \mu d\mu \nonumber   \\
           &=& \frac{(1+\sigma)^{\alpha+1}-(1+\sigma \mu^2_{\star})^{\alpha+1}}
{(1-\mu^2_{ \star })( \alpha +1 ) \sigma (1+ \sigma )^{ \alpha }}
\end{eqnarray}
Including the finite disk correction results in an increase in
$\vinf$, a decrease in $\mdot$ and a modification of
the simple scaling laws that were found in the original CAK approach 
(see Friend \& Abbott~\cite{fa86};
Pauldrach et al. 1986). The decrease of the
mass loss rate compared to the point source case is due to the 
decrease in $g_{\rm L}$ close to the star where $\dot{M}$ is 
determined. Close to the stellar photosphere the finite disk correction is 
smaller than one, viz. $D_{\rm fd}(r=R_{\star})= 1/(1+\alpha)<1$. 
The accompanying increase in $\vinf$ of typically a factor of two is due to 
two effects: (1) a reduction of $\mdot$ results in a smaller amount of material to be 
accelerated and (2) far from the photosphere, the correction factor 
$D_{\rm fd}$ becomes larger than 1. The resulting 
dependence between $\vinf$ and $\vesc$ is approximated by Friend \&Abbott (1986):

\begin{equation}
\label{eq_vinfvescabbott}
\ratio \approx 2.2 \left(\frac{\alpha}{1-\alpha}\right)
  ~\left(\frac{\vesc}{10^3~ {\rm km/s}}\right)^{0.2}
\end{equation}
They also found a relation for the modified mass loss rate:

\begin{equation}
\label{eq_mdotfa}
\mdot \approx 0.5 ~\mdot_{\rm CAK}\left( \frac{\vesc}{10^3~ {\rm km/s}} \right)^{-0.3}
\end{equation}

\begin{figure}
\resizebox{\hsize}{!}{\includegraphics{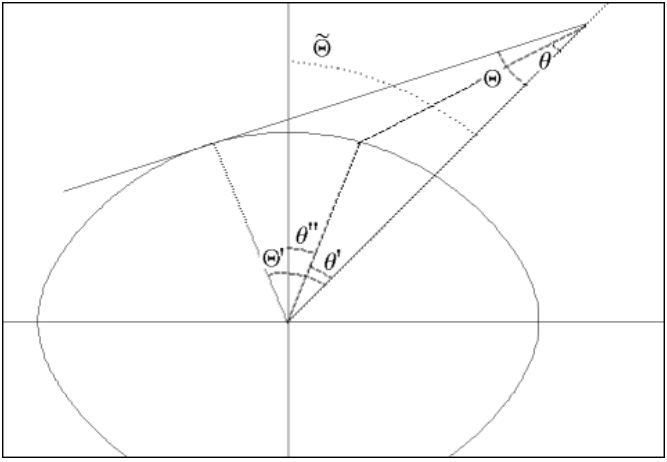}}
\caption{Geometry for the calculation of the correction factors 
$D_{\rm fd}$ and $D_{c}$. A point on the ray with co-latitude
$\tilde{\Theta}$ sees a different limb angle $\Theta$ for every $\phi$ (not
drawn; it is the angle around the line from stellar center to observer,
 not to be confused with the stellar longitude). 
 The integrals in $D_{\rm fd}$ and $D_{c}$ are over $0<\phi<2 \pi$
 and $0<\theta<\Theta$. These are rewritten to integrals
  over the star centered angles $\theta'$ ($0<\theta'<\Theta'$)
  and $\phi'=\phi$. The stellar radius and flux
 are given as a function of $\theta''$, the co-latitude, which can be 
 calculated for every $\theta'$ and $\phi$:
 $\theta''=\theta''(\theta',\phi)$.}
\label{fg_ofr}
\end{figure}

\subsection{Solution of the equation of motion for the wind of a rotating star}
\label{s_calc}

The solution of Eq.~\ref{eq_bewvglrot} gives the mass loss
rate, or rather the local mass loss rate $F_m(\theta)$, and velocity structure of
the wind. The solution is complicated however by the presence of $D_{\rm fd}$. This
is an integral of a velocity-dependent function times the surface intensity
over the visible section of the star. The appearance of the star varies
throughout the wind in
intensity distribution as well as in shape, as the star is no longer
spherical and $T_{\rm eff}$ is a function of latitude.
Since the analytic solution would be cumbersome, a numerical 
approach of the solution is chosen. This is done with the analytic
solutions of simple, i.e. the non-rotating,  
models in mind for comparison. The numerical solution of Eq.~\ref{eq_bewvglrot} is relatively
straightforward in case the function $D_{\rm fd}$ is a given function of $r$. 
Therefore we solve Eq.~\ref{eq_bewvglrot} with $D_{\rm fd}(r)$ in an 
iterative way as follows:

\begin{enumerate}

\item{A $\beta$ law for $v(r)$ is assumed, viz. $v(r)=(1-R_{\star}/r)^{0.5}$.}
\item{$D_{\rm fd}(r)$ is calculated using this velocity law.
Note that $D_{\rm fd}$ is not dependent on the actual value of $v(r)$ but
on the velocity gradient.
}
\item{Eq.~\ref{eq_bewvglrot} is solved using this correction factor $D_{\rm fd}(r)$,
obtaining a new velocity law $v(r)$.
}
\item{Step 2-3 are repeated until convergence is reached.}

\end{enumerate}

Typically three iterations are sufficient, since further iterations
changed the obtained values by less than 0.5 \%. We checked that the
solutions converged to the same value, independent of the starting
value.

\subsection{The calculation of $D_{\rm fd}$(r) and  the continuum correction factor $D_{\rm c}$}

The calculation of the correction factor $D_{\rm fd}$ from the 
velocity law (step 1. in the scheme described above) is performed by 
numerical evaluation of 
Eq.~\ref{eq_corr}. This is a non-trivial task,
since both the integrand (Eq.~\ref{eq_zeipel}) and the shape of the 
visible "disk" (Eq.~\ref{eq_rtheta}) are dependent on $r$. 

The main parameters needed to obtain $D_{\rm fd}$ for a fixed co-latitude $\tilde{\theta}$ 
are: the stellar surface $R(\theta)$ and the surface temperature
$T(\theta)$, both given as a function of the co-latitude $\theta$, 
as well as the velocity law $v(r)$. It is convenient
to change the integration variables in Eq.~\ref{eq_corr} from the wind 
centered coordinates $\theta$ and $\phi$ to the star centered coordinates
$\theta'$ and $\phi$. For the definition of these coordinates, see 
Fig.~\ref{fg_ofr}. Note that here $\phi$ is not the stellar longitude anymore, 
but the angle of rotation around the line from a point $(r,\tilde{\theta})$ in
the wind to the stellar center.    

\begin{eqnarray}
\label{eq_coint}
D_{\rm fd} = \frac{1}{N} \int_{0}^{2 \pi}\int_{0}^{\Theta(\phi)}I~ \left(
\frac{(1+\sigma) }{1+ \sigma \cos^2{\theta}} \right)^\alpha \cos{\theta}
 \sin{\theta}~d\theta d\phi \nonumber \\
 \\
 =\frac{1}{N}\int_{0}^{2 \pi}\int_{0}^{\Theta'(\phi)}I~ \left(
\frac{(1+\sigma) }{1+ \sigma \cos^2{\theta}} \right)^\alpha
\cos{\theta}
 \sin{\theta}~ \frac{d\theta}{d\theta'}~ d\theta' d\phi \nonumber
\end{eqnarray}
where $I$ is the frequency-integrated intensity

\begin{equation}
\label{eq_coint2}
I=I(\theta' ,\phi) \propto T(\theta'')^4
\end{equation}
 The angles are related to one another via 

\begin{equation}
\label{eq_angles1}
\cos \theta = \frac{r-R(\theta',\phi) \cos \theta'}
{\sqrt{(r-R(\theta',\phi) \cos \theta')^2+(R(\theta',\phi) \sin
    \theta')^2} }
\end{equation}
with the stellar surface given by $R(\theta',\phi)=R(\theta'')$, and
\begin{equation}
\label{eq_angles2}
\cos \theta''=\sin \theta' cos \phi \sin \omega+\cos \theta' \cos \omega
\end{equation}
Equation~\ref{eq_coint} can be integrated, considering that the limb 
angle $\Theta'(\phi)$ can be determined numerically, yielding the finite disk 
correction for the line acceleration of an arbitrary shaped star.

The correction factor $D_{\rm c}$ for the electron acceleration 
 (defined in Eq.~\ref{eq_dc})
similarly becomes:

\begin{equation}
\label{eq_cc}
D_{c}=\frac{1}{N} \int_{0}^{2 \pi}\int_{0}^{\Theta'} I~ 
\cos{\theta}
 \sin{\theta}~ \frac{d\theta}{d\theta'}~ d\theta' d\phi
\end{equation}
This correction is also included in the solution of
the equation of motion.


\begin{table*}
\label{tb_res1}
\caption[]{Properties of a typical rotating B[e] star model $^1$ } 

\begin{tabular}{c|ccc|ccc|ccc|ccc|cccc} 
  & & $\theta=0~~$ & & & $\theta=\pi/8$ & & & $\theta=\pi/4$ & & & $\theta=3 \pi/8$ 
  & & & $\theta=\pi/2$ & \\
$\omega$ &  $\vinf$ $^2$ & $F_m$ $^3$
& $\beta$ $^4$ & $\vinf$ & $F_m$&$\beta$ & $\vinf$ & $F_m$&$\beta$
& $\vinf$ & $F_m$&$\beta$ & $\vinf$ & $F_m$ & $\beta$ & \\
\hline
0 &  1.40 & 2.35 & 0.63 & & & & & & & & & & & & & \\
.3 &  1.43 & 2.67 & 0.63 & 1.40 & 2.64 & 0.63 & 1.35 & 2.53 & 0.63 &
 1.30 & 2.42 & 0.63 & 1.29 & 2.36 & 0.64 & \\
.4 &  1.44 & 2.95 & 0.63 & 1.40 & 2.88 & 0.62 & 1.31 & 2.70 & 0.62 &
 1.23 & 2.46 & 0.63 & 1.20 & 2.34 & 0.64 & \\
.5 &  1.46 & 3.33 & 0.63 & 1.40 & 3.23 & 0.62 & 1.27 & 2.95 & 0.62 &
 1.14 & 2.53 & 0.64 & 1.11 & 2.27 & 0.66 & \\
.6 &  1.48 & 3.83 & 0.63 & 1.40 & 3.70 & 0.62 & 1.21 & 3.29 & 0.60 &
 1.04 & 2.61 & 0.64 & 1.01 & 2.11 & 0.70 & \\
\hline 
\end{tabular}

(1) The adopted stellar parameters are:\\
 $\teff=20000$ K, $L_{\star}=10^{5.5} \lsun$, $\mster=40 \msun$,
$R_{\star}= 47 \rsun$, $\alpha=0.565$, $k=0.32$, $\delta=0.02$, 
solar abundances.\\
(2): the terminal velocity $\vinf$ is in $10^3$ km/s \\
(3): the mass flux $F_m$ in $10^{-6}$ \msunyr \\
(4):  the velocity law parameter $\beta$ is obtained from a
nonlinear fit of $v(r)$ from 1.1 $R_*$ to 10 $R_*$.

\end{table*}

\begin{table*}
\label{tb_res2}
\caption[]{Properties of rotating B[e]$^1$ models: with different values of
 $L_{\star}$ }

\begin{tabular}{c|ccccccc}
$L_*$ & $\omega$ & $v_{inf}^{pole}$ & $F_m^{pole}$ & $v_{inf}^{eq}$
 & $F_m^{eq}$  &
$\frac{\rho_{\rm eq}}{\rho_{\rm pole}}$ & $\mdot$  \\
$L_{\odot}$ &   & $10^3$ \kms  & $10^{-8}$ \msunyr & $10^3$ \kms & $10^{-8}$
\msunyr &  & $10^{-8}$ \msunyr \\
\hline
$10^{4.5}$  &   0 & 2.84 & 3.01 &      &      & 1    & 3.01 \\ 
  & 0.3 & 2.91 & 3.37 & 2.60 & 3.04 & 1.01 & 3.18 \\
  & 0.4 & 2.97 & 3.67 & 2.42 & 3.03 & 1.01 & 3.32 \\
  & 0.5 & 3.04 & 4.08 & 2.22 & 2.98 & 1.00 & 3.50 \\
  & 0.6 & 3.11 & 4.62 & 2.00 & 2.82 & 0.95 & 3.71 \\
\hline
\end{tabular}\\

\vspace{0.3cm}
\begin{tabular}{c|ccccccc}
$L_*$ & $\omega$ & $v_{inf}^{pole}$ & $F_m^{pole}$ & $v_{inf}^{eq}$
 & $F_m^{eq}$  &
$\frac{\rho_{\rm eq}}{\rho_{\rm pole}}$ & $\mdot$  \\
$L_{\odot}$ &   & $10^3$ \kms  & $10^{-7}$ \msunyr & $10^3$ \kms & $10^{-7}$
\msunyr &  & $10^{-7}$ \msunyr \\
\hline
$10^{5.0}$ & 0 & 2.06 & 2.53 &      &      & 1    & 2.53\\
 & 0.3 & 2.11 & 2.84 & 1.88 & 2.55 & 1.00 & 2.67\\
 & 0.4 & 2.15 & 3.11 & 1.75 & 2.54 & 1.00 & 2.79\\
 & 0.5 & 2.19 & 3.46 & 1.61 & 2.49 & 0.98 & 2.94\\
 & 0.6 & 2.24 & 3.93 & 1.46 & 2.35 & 0.91 & 3.15\\
\hline
\end{tabular}

\vspace{.3cm}
\begin{tabular}{c|ccccccc}
$L_*$ & $\omega$ & $v_{inf}^{pole}$ & $F_m^{pole}$ & $v_{inf}^{eq}$
 & $F_m^{eq}$  &
$\frac{\rho_{\rm eq}}{\rho_{\rm pole}}$ & $\mdot$  \\
$L_{\odot}$ &   & $10^3$ \kms  & $10^{-6}$ \msunyr & $10^3$ \kms & $10^{-6}$
\msunyr &  & $10^{-6}$ \msunyr \\
\hline
$10^{5.5}$ &  0  & 1.40 & 2.35 &      &      & 1    & 2.35 \\
 & 0.3 & 1.43 & 2.67 & 1.29 & 2.36 & 0.98 & 2.47 \\
 & 0.4 & 1.44 & 2.95 & 1.20 & 2.34 & 0.95 & 2.56 \\
 & 0.5 & 1.46 & 3.33 & 1.11 & 2.27 & 0.90 & 2.71 \\
 & 0.6 & 1.48 & 3.83 & 1.01 & 2.11 & 0.80 & 2.89 \\
\hline
\end{tabular}

\vspace{.3cm}
\begin{tabular}{c|ccccccc}
$L_*$ & $\omega$ & $v_{inf}^{pole}$ & $F_m^{pole}$ & $v_{inf}^{eq}$
 & $F_m^{eq}$  &
$\frac{\rho_{\rm eq}}{\rho_{\rm pole}}$ & $\mdot$  \\
$L_{\odot}$ &   & $10^3$ \kms  & $10^{-5}$ \msunyr & $10^3$ \kms & $10^{-5}$
\msunyr &  & $10^{-5}$ \msunyr \\
\hline
$10^{6.0}$ & 0 & 0.65 & 3.93 &      &      & 1    & 3.93\\
 & 0.3 & 0.62 & 4.97 & 0.60 & 3.80 & 0.78 & 4.25\\
 & 0.4 & 0.59 & 6.00 & 0.58 & 3.61 & 0.62 & 4.54\\
 & 0.5 & 0.55 & 7.79 & 0.56 & 3.24 & 0.41 & 5.01\\
 & 0.6 & 0.49 & 11.1 & 0.56 & 2.64 & 0.21 & 5.76\\
\hline
\end{tabular}

(1): The adopted stellar parameters are: $M_{\star}=40 M_{\odot}$; $R_{\star}=47 R_{\odot}$; $\teff = 20000$ K. 
\end{table*}


\begin{table*}
\label{tb_res3}
\caption{Properties of rotating B[e] $^1$ models: with different values of 
$M_{\star}$ } 
\begin{tabular}{c|ccccccc}
$M_*$ & $\omega$ & $v_{inf}^{pole}$ & $F_m^{pole}$ & $v_{inf}^{eq}$
 & $F_m^{eq}$  &
$\frac{\rho_{\rm eq}}{\rho_{\rm pole}}$ & $\mdot$  \\
$M_{\odot}$ &   & $10^3$ \kms  & $10^{-6}$ \msunyr & $10^3$ \kms & $10^{-6}$
\msunyr &  & $10^{-6}$ \msunyr \\
\hline
20 & 0   & 0.80 & 5.66 &      &      & 1    & 5.66 \\
 & 0.3 & 0.80 & 6.64 & 0.73 & 5.67 & 0.94 & 5.69 \\
 & 0.4 & 0.80 & 7.49 & 0.68 & 5.58 & 0.88 & 6.00 \\
 & 0.5 & 0.79 & 8.74 & 0.64 & 5.32 & 0.75 & 6.38 \\
 & 0.6 & 0.79 & 10.5 & 0.60 & 4.76 & 0.53 & 6.81 \\
\hline
\end{tabular}

\vspace{0.3cm}
\begin{tabular}{c|ccccccc}
$M_*$ & $\omega$ & $v_{inf}^{pole}$ & $F_m^{pole}$ & $v_{inf}^{eq}$
 & $F_m^{eq}$  &
$\frac{\rho_{\rm eq}}{\rho_{\rm pole}}$ & $\mdot$  \\
$M_{\odot}$ &   & $10^3$ \kms  & $10^{-6}$ \msunyr & $10^3$ \kms & $10^{-6}$
\msunyr &  & $10^{-6}$ \msunyr \\
\hline
60 & 0 & 1.82 & 1.55 &      &      & 1    & 1.55\\
 & 0.3 & 1.85 & 1.76 & 1.66 & 1.56 & 0.99 & 1.55\\
 & 0.4 & 1.89 & 1.92 & 1.56 & 1.55 & 0.98 & 1.59\\
 & 0.5 & 1.91 & 2.16 & 1.43 & 1.51 & 0.93 & 1.68\\
 & 0.6 & 1.95 & 2.46 & 1.31 & 1.41 & 0.85 & 1.75\\
\hline
\end{tabular}

(1): The adopted stellar parameters are: $L_{\star}=10^{5.5} L_{\odot}$; $R_{\star}=47 R_{\odot}$; $\teff = 20000$ K. 
\end{table*}


\subsection{Solving the equation of motion}

To solve the equation of motion it is neccesary to determine the conditions
at the critical point: $r_c$, $v_c$ and $v'_c$, or, equivalently, $r_c$,
$v_c$ and $v_0=v(R_{\star})$, as well as the local mass loss rate
$F_m$. Apart from Eqs.~\ref{eq_sing} and~\ref{eq_regu} and the equation of
motion (\ref{eq_bewvglrot}) itself the additional constraint 
 for the continuum optical depth $\tau_{\rm e}$ at the stellar radius
  is neccesary (Pauldrach et al. 1986) to uniquely determine these 
  quantities:

\begin{equation}
\label{eq_tau2/3}
\tau_{\rm e}(\theta) = \int_{R_{\star}}^{\infty} \rho \sigma_{\rm e} dr = \frac{2}{3}
\end{equation}
If the condition for $v_0$ is used one may adopt any small value, 
say $v_0=0.1$ km/s, since
even a large error in $v_0$ will only yield a relatively small error in $\vinf$ and
$F_m$. This is due to the fact that at the surface of the star $v(r)$ and thus also $\rho$
vary exponentially with scale height
$H=\frac{a^2 R_*^2}{G M_{\star}^{\rm eff}} << R_{\star}$. 
This means that the radius where Eq.~\ref{eq_tau2/3} is fulfilled is
very close to $R_{\star}$. This has been checked a posteriori.
In our calculations we have adopted
this boundary condition for $v_0$ in the
solution of the full momentum equation.

A first guess for the local mass loss rate obtained from the CAK solution 
is used to integrate the equation of motion from the stellar surface
outward. This gives an approximate value for the 
critical point where $r^2 v \frac{dv}{dr}=0$
if $F_m ({\rm guess})< F_m({\rm solution})$. 
This approximate value for the critical point can then be used to 
consecutively improve the guessed value of $F_m$ using Eqs.~\ref{eq_sing} and~\ref{eq_regu}.
This iterative procedure is terminated when the local mass loss rate converges. The  
calculation is then completed by integrating 
past the critical point on the solution which extends to infinity
(see CAK or Abbott 1980 for a description of the
topology of the solutions). To deal with a numerical singularity at the sonic 
point the equations are solved for $r(v)$ rather than for $v(r)$.


\section{Application to B[e] winds}
\label{s_B[e]}

\subsection{A typical B[e] supergiant}

\begin{figure}
\resizebox{\hsize}{!}{\includegraphics{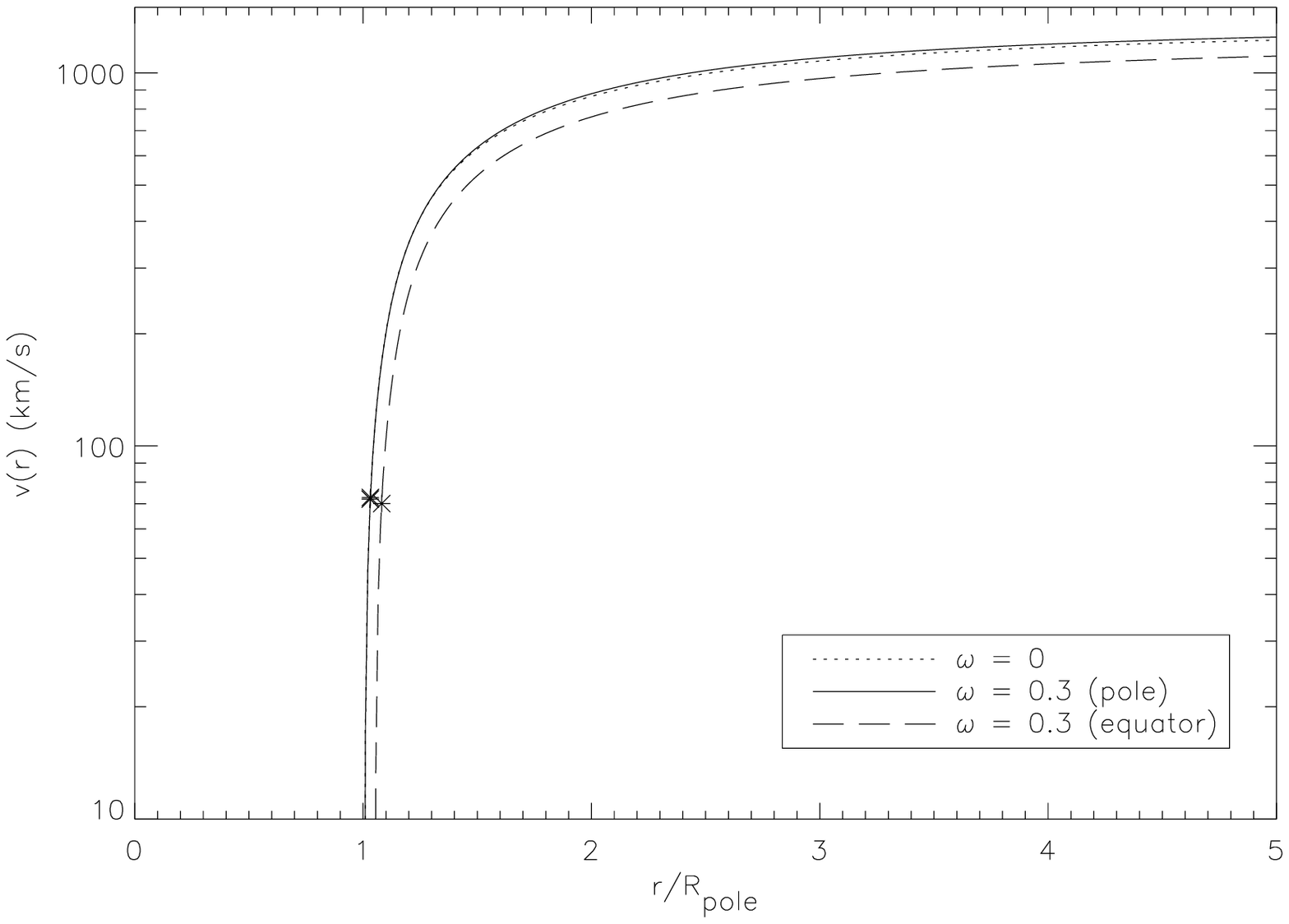}}
\resizebox{\hsize}{!}{\includegraphics{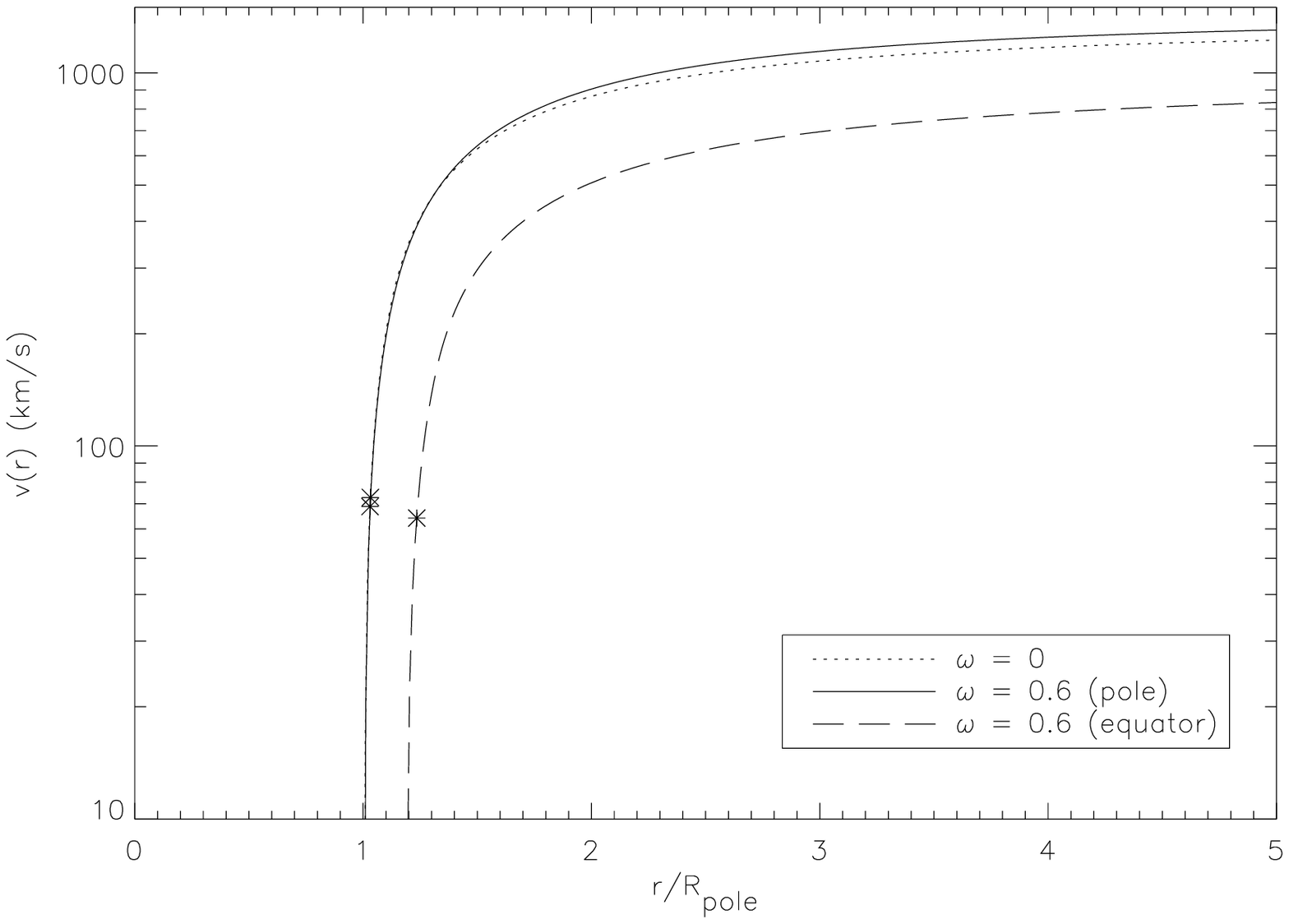}}
\caption{The velocity law $v(r)$ at the pole and equator of B[e] wind models
 for different rotation
 rates. Top panel: $\omega=0.3$, bottom panel: $\omega=0.6$.
 The crosses indicate the location of the  critical point
  of the wind. Note that the equatorial curves in the rotating models start 
  at a radius $r/R_{\rm pole}>1$ due to the oblateness of the star (see text for stellar
  parameters).}
\label{fg_vr}
\end{figure}

\begin{figure}
\resizebox{\hsize}{!}{\includegraphics{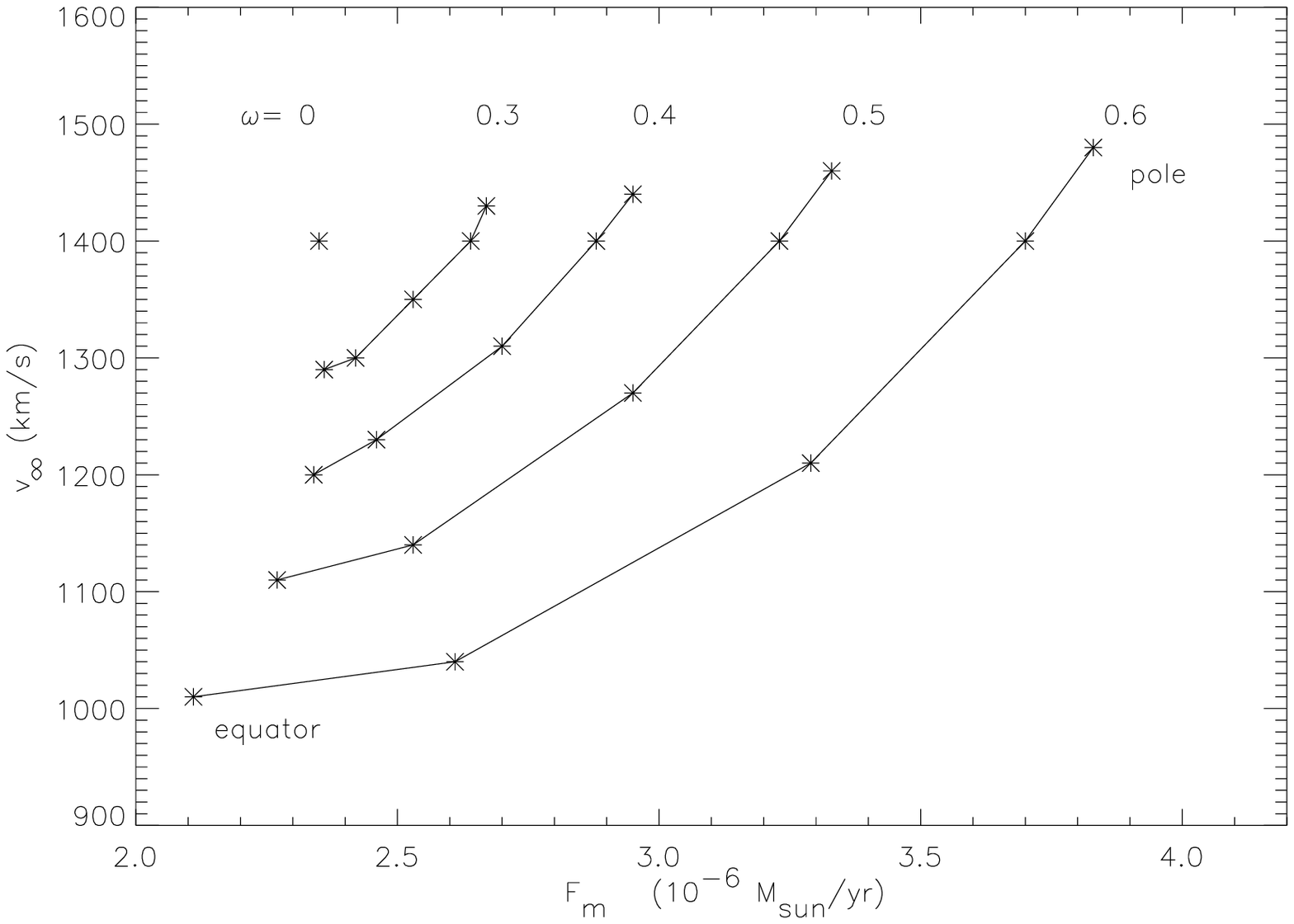}}
\caption{The latitude dependence of the mass-loss rate and $\vinf$ of a
typical B[e] star, for various rotational speeds.
(See text for stellar parameters).}
\label{fg_vmp}
\end{figure}

We have calculated models for rotating B[e] stars using the above described
method. For the first models we adopted the following stellar
parameters, which are typical for B[e] stars:
 $L_{\star}=10^{5.5} \lsun$,
$\mster=40 \msun$,
$R_{\star, \rm pole}= 47 \rsun$, $\teff \simeq 20~000$ and solar abundances.
The effective temperature is defined as $\teff=(L_*/\sigma_B S)^{0.25}$
where $S$ is the total surface of the distorted star.
The following force multiplier parameters from Pauldrach et al. 
~(\cite{ppk86}) were adopted to
describe the line acceleration: $\alpha=0.565$, $k=0.32$, $\delta=0.02$.
An overview of the results for this generic B[e] star rotating at
$\omega$ = 0, 0.3, 0.4, 0.5 and 0.6  times the critical
 rotation speed is given in Table 1.
The calculation of models with higher values of $\omega$ runs into  numerical
problems, because in the iteration,  the approximate critical point
can be  below the sonic point.
The table shows for
 different co-latitudes $\theta$ and different rotation velocities $\omega$
 the values of $\vinf$, the local mass loss rate 
$F_m(\theta)$ and the value of the
 velocity law parameter $\beta$, obtained by fitting the calculated
velocity structure from
 $1.1~ R_{\star}$ to $10~ R_{\star}$ to a
  $\beta$-law. We see that the value of $\beta$ does not change much.

Fig.~\ref{fg_vr} shows the velocity structure of the wind and the location of
the critical point for $\omega$ = 0, 0.3 and 0.6. Note that the critical point is very 
close the star. This is true for all finite disk corrected solutions. 
Figure~\ref{fg_vmp} shows the behaviour of $\vinf$ versus
$F_m$ for various rotational speeds and latitudes. 
As expected, $\vinf$ at the equator is smaller than at the pole because
$\vinf$ scales roughly with $\vesc$ and due to the 
smaller value of $\vesc$ at the equator.

\begin{figure}
\resizebox{\hsize}{!}{\includegraphics{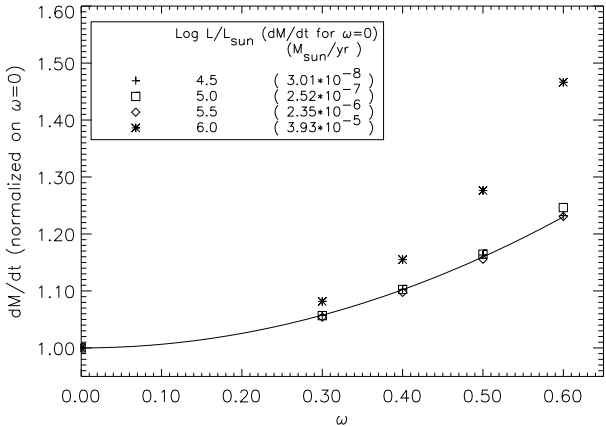}}
\caption{Mass loss rates for different rotation speeds.
The drawn line is given by: $1+0.64 \omega^2$}
\label{fg_mp}
\end{figure}

Figure \ref{fg_mp} shows the latitude dependence of the local
mass-loss rate and of the terminal velocity from equator to pole,
for different rotational speeds, indicated by $\omega$.
At the pole ($\theta=0$) $\vinf$ is about constant for various
rotational speeds.  
The mass-loss rate at the equator decreases slightly with increasing
rotation. 
The increase of the mass loss rate at the equator as calculated
by Pauldrach et al. (\cite{ppk86}) and Friend \& Abbott (\cite{fa86}) 
is offset by a decrease in mass loss rate due to the
smaller radiative flux at the equator due to the von Zeipel effect. 
The mass loss rate at the pole increases strongly with increasing
rotation rate 
due to the increase in luminosity at the pole. This is because the
total luminosity of the star  
is fixed and the smaller luminosity at the equator must therefore be 
compensated by a higher luminosity at the pole. 

\subsection{The overall density properties}

The overall effect of rotation is to increase the
mass loss rate of a star. This can be seen in Fig.~\ref{fg_mp} where the
overall mass-loss rate (i.e. $F_m$ integrated over the stellar surface)
is plotted vs. the rotational velocity. 
The mass-loss rate of the models with 
$4.5 \le \log L_*/L_{\odot} \le 5.5$ vary with $\omega$ as 
$1+0.64 \omega^2$.
 
\begin{figure}
\resizebox{\hsize}{!}{\includegraphics{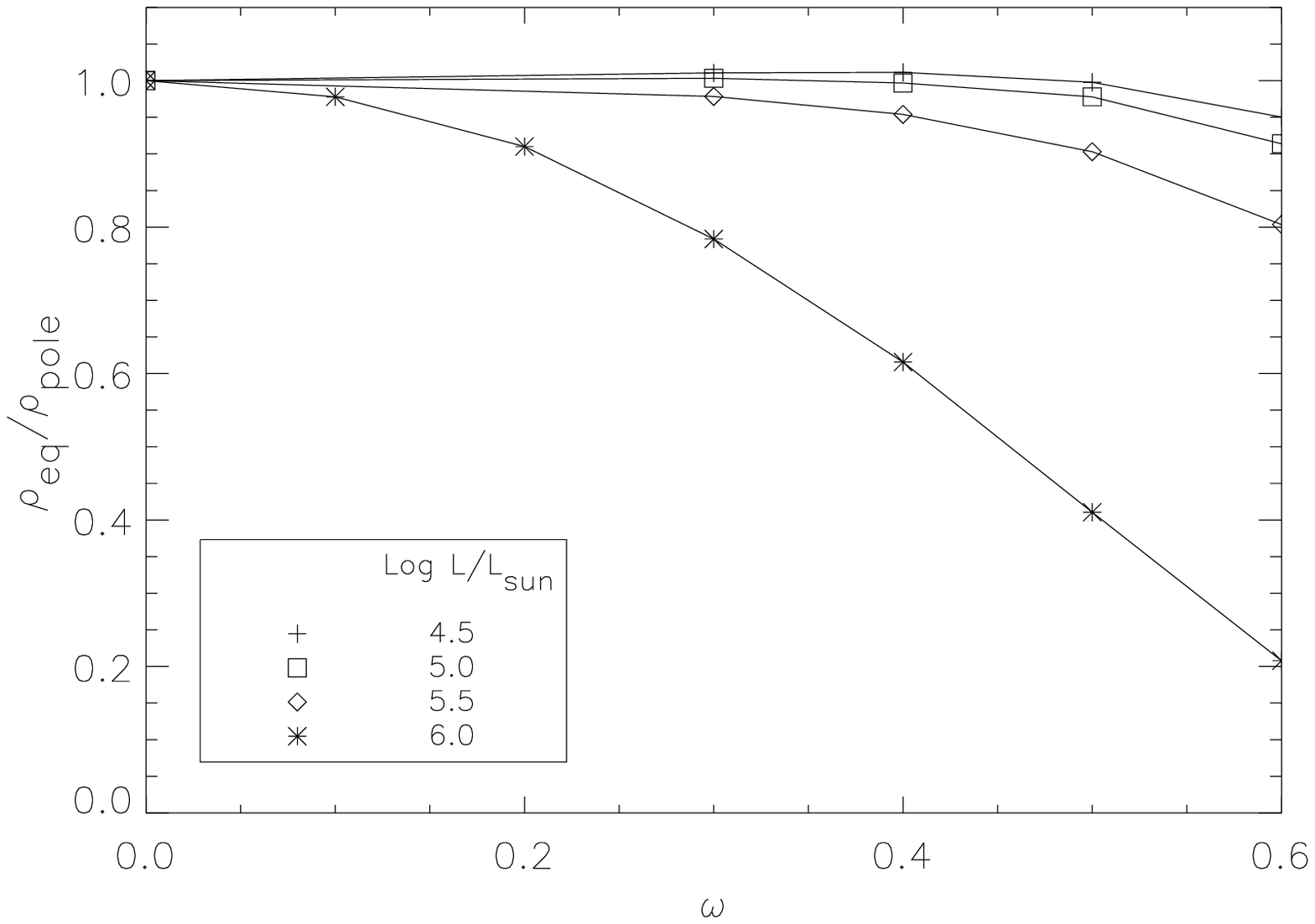}}
\caption{The density contrast
$\rho^{\rm eq}/\rho^{\rm pole}$ for various values of $\log{L_{\star}/\lsun}$. 
}
\label{fg_ldens}
\end{figure}

Figure~\ref{fg_ldens} shows the resulting density contrast (far
from the star)
 between pole and equator
( $\rho_{\rm eq} /\rho_{\rm pole}=F_m^{\rm eq} v_{\infty}^{\rm pole}/
F_m^{\rm pole} v_{\infty}^{\rm eq}$). The densities at pole and equator are
essentially the same: the smaller $\vinf$ at the equator is offset by the
larger mass loss rate $F_m$ at the pole.

\subsection{Varying $L_{\star}$}

\begin{figure}
\label{fg_dc}
\resizebox{!}{7.0cm}{\includegraphics{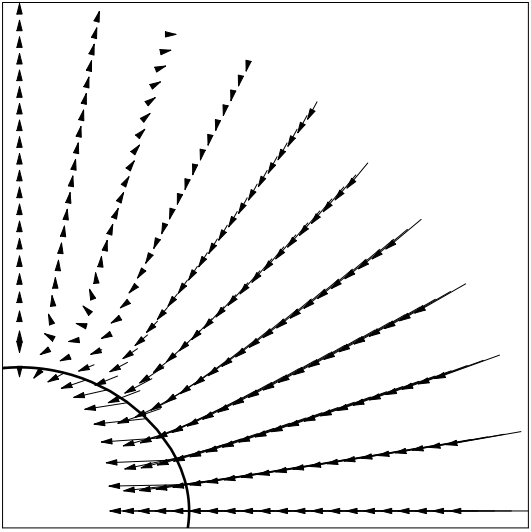}}
\caption{Vector plot of the gravity ($F_{\rm g}$) and the continuum
radiation force ($F_{\rm e}$).
The vectors indicate strength and direction of $r^2 (F_{\rm g}+ F_{\rm e})$ on an arbitrary
scale for a star with $\Gamma_e=0.76$ and $\omega=0.6$. Note that line
forces and centrifugal forces are not included.}
\end{figure}

We have also calculated models for different luminosities (see Table 2),
and different stellar masses (see Table 3).
The mass-loss rate for the different luminosities is
also plotted in Fig.~\ref{fg_mp}.
The density contrast for various luminosities is plotted
in Fig.~\ref{fg_ldens}. We see that the density at the pole compared 
to the equator increases for increasing 
luminosity and the effect on the mass-loss rate is only visible for the highest
luminosity. This is an effect of the continuum acceleration $D_{\rm c}$.
Fig. 7 illustrates the effect of the continuum acceleration on
 the forces in the wind for a star with high luminosity
 ($\Gamma_{e}=0.76$).
The net effect of the continuum radiation pressure and the
gravity, i.e. $G M_*(1-\Gamma_e')/r^2$, 
is an inward force in the equatorial region and an outward force
in the polar region where the flux is higher.

 For small luminosity the variation in $\Gamma_{\rm e}'$ through 
$D_{\rm c}$ is 
 not important because 
$\Gamma_{\rm e}$ itself is small. For high luminosity $L_*/(G \mster)$ increases and 
thus the effect 
of $D_c$ on $\Gamma_{\rm e}'$ becomes noticeable (this is with all other
properties of the star fixed). In Eq.~\ref{eq_vinfsimple} we see that a
larger value of $D_c$ at the pole causes a relatively larger decrease
(through $\Gamma_{\rm e}'$) in $\vesc$ and thus in $\vinf$, whereas the mass loss 
rate at the equator will increase less because of the smaller $D_{\rm c}$.
This is only a qualitative explanation and the actual dependence of
$\rho_{\rm eq}/\rho_{\rm pole}$ does not follow the relation suggested by
Eqs. 29 and 31. We have found that the actual values of 
$\rho_{\rm eq}/\rho_{\rm pole}$ could be approximated quite well (within
10 percent) with the following relation:

\begin{equation}
\label{eq_lrho}
\rho_{\rm eq}/\rho_{\rm pole} \approx \left\{ \frac{1-\Gamma_{\rm
e}'(\theta=0,r/R_{\rm eq}=1)}{1-\Gamma_{\rm e}'(\theta=\pi/2,r/R_{\rm
eq}=r_{\rm min})} \right\}^{1.5}
\end{equation}
where the $r_{\rm min} \approx 1.2$ is the radius where $D_{\rm c}$
reaches its minimum value.

So we see that although rotation alone modifies the wind structure 
considerably, it cannot be responsible for the density contrast
between the equatorial and polar wind observed in B[e] stars,
in case only radial effects are considered. We conclude that at least
one other physical effect must be responsible for the formation of
disks of rotating B[e] stars. The two most likely additional effects
are the bi-stability  and the flow of the wind
material towards the equator.


\section{Rotationally induced bi-stability models}
\label{s_bs}

In the section above we have shown that rotation alone cannot
explain the observed large density contrast between the poles and the
equator of B[e] stars. One of the possible additional mechanisms
to enhance this contrast is the bi-stability jump.

To investigate whether the bi-stability jump observed in  normal
B-supergiants can explain the formation of disks around
B[e] stars, we
have calculated bi-stable wind models for a fast rotating
typical B[e] supergiant. 
Spherical, NLTE wind models for normal B supergiants have been 
calculated by Vink et al. (1999). From these models CAK-like force multiplier
parameters have been derived using a Monte-Carlo method to simulate photon-gas
interactions. These models show the occurence of a bi-stability jump
around $\teff \simeq 25 000$ K. 

For two models on either side of the bi-stability jump ($\teff$ = 17 500 and
30 000 K), the CAK parameters were determined by
fitting a power law from about the sonic point to about $0.5 \vinf$ (from 
$t=10^{-2}$ to $t=10^{-4}$ in optical depth, $t$ is defined in Eq.~\ref{eq_t})
to the calculated force multiplier values.
The resulting values for $k$, $\alpha$ and $\delta$ are
listed in Table~\ref{tb_ka}. 
Note that the $\delta$ parameter is taken equal to zero as it was not
possible to extract it explicitly from the models. Its effect is
hidden in the constant $k$. 
As expected, the values for the force multiplier parameters are quite different for the two cases, because the ionization has changed dramatically
over the bi-stability jump.

\begin{table}
\caption{The force multiplier parameters for a bi-stable wind}
\begin{tabular}{cccc}
  \teff & $k$ & $\alpha$ & $\delta$\\
\noalign{\smallskip}
\hline
17500 K & 0.57 & 0.45 & 0.0\\
30000 K & 0.06 & 0.65 & 0.0\\
\noalign{\smallskip}
\hline
\end{tabular}
\label{tb_ka}
\end{table}

The predicted bi-stability jump occurs around $\teff \simeq 25$ kK. Therefore we calculated
a model for a rotating B[e] star with
the following properties: $\teff=25 000$ K, $L_{\star}=10^5 \lsun$,
$M_{\star}=17.5 \msun$ and solar abundances.
This implies that the pole of the rapidly rotating star will be hotter
than 25 kK and the equator will be cooler than 25 kK.
 
For the pole, the force multipliers $k$ and $\alpha$ from the hot
Monte Carlo model of 30 000 K were used, whereas
for the equator, $k$ and $\alpha$ from the cool model of 17 500 K were
used. 

\begin{figure}
\resizebox{\hsize}{!}{\includegraphics{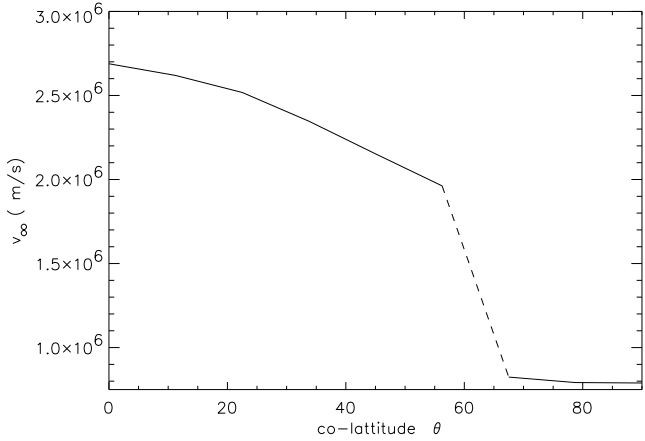}}
\caption{The terminal velocity $\vinf$ for a star with a bi-stability jump
as a function of co-latitude. The dashed part indicates the location
of the bi-stability jump.
Notice the drastic drop by about a factor 3 
around $\theta$=60$^0$, due to the bi-stability jump.
(See text for the adopted stellar parameters)}
\label{fg_vtheta}
\end{figure}
\begin{figure}
\resizebox{\hsize}{!}{\includegraphics{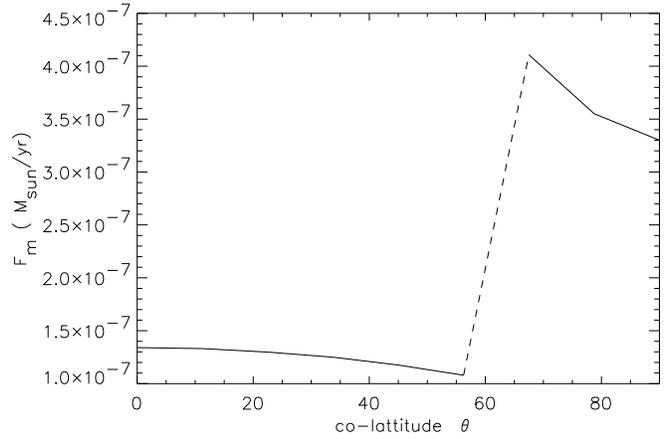}}
\caption{The mass loss rate $F_m$ of a star with a bi-stability jump
as a function of co-latitude. The dashed part indicates the location
of the bi-stability jump. Notice the steep jump in mass loss rate near 
$\theta = 60^0$.
(See text for the adopted stellar parameters)
}
\label{fg_mdtheta}
\end{figure}

The resulting $\vinf(\theta)$ and mass loss rates $F_m(\theta)$ are 
plotted in Figs.~\ref{fg_vtheta} and 
\ref{fg_mdtheta} for $\omega$ = 0.6. Clearly visible is 
the drastic decrease in
$\vinf$ and the drastic increase of $F_m$ towards the equator.
This occurs around a co-latitude $\theta$ (angle
from the pole) of about 60 degrees, where $\teff \simeq 25~000$ K. (The precise location of the jump depends on
the effective temperature of the star).
Since both $F_m$ and $\vinf$ have a equator/pole contrast of about 
a factor of three, the resulting density contrast is about a factor of 10
in a disklike region with a half opening angle of 30 degrees. 

Our calculations show that 
the density contrast in the wind between pole and equator 
due to the bi-stability jump is significantly larger than without 
the bi-stability jump. Yet, the calculated value of about a factor 10 
is not sufficient to explain the observed density contrast of
a factor $ \sim 100$. So another mechanism is needed to enhance
the density contrast even further.
This is most likely the
wind compression mechanism (Bjorkman \&~ Cassinelli 1993; Bjorkman 1998).


\section{Summary and discussion}
\label{s_concl}

We have modified the line driven wind theory for rotating stars by
including the effects of oblateness and gravity darkening as well
as rotational terms in the equation of motion. 
We considered a
sectorial wind, i.e. we neglected the effect
of the motion of the wind towards the equator or towards the pole.
This assumption is justified close to the stellar surface below the
critical point of the momentum equation. This implies that our method
will predict about the correct distibution of mass-loss rates from the star
as a function of stellar latitude, but it may not be accurate enough 
to predict the velocity and density distribution further away from the
star, if motions in the $\theta$ direction become important.

The equation of motion was solved using an iterative numerical scheme,
that includes the conservation of angular momentum and
the correction factors to the radiative acceleration
by lines and by the continuum, due to the non-spherical shape
of the star and due to the latitude dependence of the radiative flux.
This method was applied to study its possible effects on the formation of 
disks around fast rotating B[e] supergiants.

The models with constant force multiplier parameters $k$, $\alpha$ and $\delta$
show a decrease of {\it both} the mass loss rate and the terminal velocity from
pole to equator. This is mainly due to two effects: the reduction of the escape
velocity from pole to equator, resulting in a higher terminal velocity
at the pole, and the reduction of the radiative flux
from the pole to the equator due to gravity darkening, which results in a
decrease in mass loss rate at the equator.

For a star with a fixed luminosity and a fixed polar radius, 
the temperature at the pole
increases and the temperature at the equator decreases with increasing
rotation rate. The terminal velocity of the wind
from the poles is almost independent of $v_{\rm rot}$
 but at the equator $\vinf$ decreases with increasing rotation rate.
  The mass-loss rate at the pole
increases with increasing $v_{\rm rot}$, due to the increase in $\teff$
at the poles, but the mass loss rate from the equator is almost
independent of the rotation. The combination of these effects alone
produce a density contrast between the polar and the equatorial
wind of a factor $\rho_{\rm eq}/\rho_{\rm pole} \simeq 1$,
 except for fast rotating stars with high luminosity (in which case 
 $\rho_{\rm eq}/\rho_{\rm pole} < 1$).

Our results agree quantitatively with those obtained by Maeder (\cite{mae})
for the latitudinal dependence of $F_m$. Our predictions of the
global mass loss rates are also in general agreement
with those of Maeder for models with luminosities
far from the Eddington limit. For high luminosities we find a strong
polar outflow, whereas Maeder's  models predict an enhanced equatorial
outflow on the basis of the original CAK scaling laws.
This difference is due to the
fact that we have included an improved description of the radiation
pressure in a rotating star.

The difference between the polar wind and the equatorial wind
is strongly enhanced when the bi-stability of 
radiation driven winds between $\teff \simeq 20~000$ and 30~000 K 
is taken into account.
In this case the force multipliers $k$ and $\alpha$ of a rapidly rotating
star change drastically with stellar latitude if the pole is hotter
than 25~000 K and the equator is cooler. In \S~\ref{s_bs} we have applied the newly
calculated force multipliers from Vink et al. (1999) 
above and below the bi-stability jump
 to the models of rotating B[e] supergiants.
One might argue that it is not allowed to apply 
the force multipliers of (bi-stable) spherical wind models 
to aspherical winds of rotating stars. 
However, we have shown that a disk formed through bi-stability 
will be sufficiently thick (typically about 30 degrees for a star with an effective
temperature equal to about 25~000) to make these
spherical models reasonable approximations for the conditions in the
wind close to the star, where the mass loss rate is determined.
We find that rotationally induced bi-stability models of B[e] stars
reach a density contrast of about a factor 10 between the dense
equatorial wind and the less dense polar wind. This is less than
the factor $10^2$ that is observed (see e.g. Bjorkman~\cite{bj98}).

The extra density increase is most likely due to the wind compression.
The gas that leaves the photosphere from a rotating wind, will
follow an orbit in a tilted plane defined by the local rotation
vector and the center of the star. If the rotational velocity is large
or the wind velocity is small, this orbit will cross the equatorial 
plane where the streamlines of the wind from different
stellar latitudes cross. The resulting shock will compress the gas
into a thin outflowing disk with an opening angle of only a few
degrees (Bjorkman \& Cassinelli 1993; $ISW$ Chapt 11).
We have not considered this wind compression in our model. However,
it is likely to occur in rotationally induced bi-stable winds, because
the wind velocity in the equatorial plane is about a factor three
smaller than from the poles, thus facillitating the wind compression.

Owocki \& Cranmer (1994) have argued that the flow of the wind towards the 
equatorial plane, predicted in the WCD-theory, may be offset by a
$\theta$-component of the line acceleration towards the polar regions.
This ``disk inhibition'' mechanism operates in wind models with
constant force multipliers $k$ and $\alpha$. However, it is not clear
that this mechanism is sufficiently strong to overcome the 
wind compression in the RIB-model. This is because $k$ and $\alpha$
change with latitude in the RIB-model and there is a strong
density gradient in the wind from the equator to the poles
(Owocki et al. 1998, Puls et al. 1999).

The combination of the RIB and the WCD mechanism may offer the
best possibility for explaining the disks of B[e]-supergiants.
Whether the compression is strong enough to explain the observed
high density contrast between the polar and the equatorial wind,
remains to be seen. It can be calculated by combining the
solutions of the wind momentum equation of rotating oblate
stars with gravity darkening (derived in this paper), and the  
calculation of the resulting trajectories of the wind.
The combination of the rotation induced bi-stability model and the
wind compressed disk model is promising for explaining the disks
of B[e] stars because the RIB-mechanism explains the increased
mass loss and the small velocity from the equatorial regions 
and the WCD-mechanism explains the strong  compression of the disk.


\begin{thebibliography}{}


\bibitem[1980]{ab80}
        Abbott D.C., 1980, ApJ 242, 1183
\bibitem[1982]{ab82}
        Abbott D.C., 1982, ApJ 259, 282
\bibitem[1998]{bj98}
        Bjorkman J.E., 1998, in {\it B[e] stars}, eds A.M. Hubert, C. Jaschek
        Kluwer: Dordrecht, ASSL 233, p 189 
\bibitem[1993]{bc93}
        Bjorkman J.E., Cassinelli J.P., 1993, ApJ 409, 429
\bibitem{}
        Cassinelli 1998 in {\it B[e] stars}, eds A.M. Hubert, C. Jaschek
        Kluwer: Dordrecht, ASSL 233, p 177 
\bibitem[1974]{ca74}
        Castor J.I., 1974, MNRAS 169, 279
\bibitem[1975]{cak}
       Castor J.I., Abbott D.C., Klein R.I., 1975, ApJ 195, 157
\bibitem[1994]{co94}
       Cranmer S.R., Owocki S.P., 1994, ApJ 440, 308
\bibitem{}
        Feldmeier A., 1999 in {\it Variable and non-spherical stellar winds in
        luminous hot stars}, Lecture Notes in Physics, eds. B. Wolf et al.
        Springer: Heidelberg, p 285.
\bibitem[1986]{fa86}
        Friend D.B., Abbott D.C., 1986, ApJ 311, 701
\bibitem[1995]{ga95}
        Gayley K.G., 1995, ApJ 454, 410
\bibitem[1989]{kppa89}
        Kudritzki R.P., Pauldrach A., Puls J., Abbott D.C., 1989, A\&A 219, 205
\bibitem[1999]{lc99}
        Lamers H.J.G.L.M., Cassinelli J.P., 1999, Introduction to Stellar Winds,
        Cambridge Univ. Press. ($ISW$)
\bibitem[1991]{lp91}
        Lamers H.J.G.L.M., Pauldrach A., 1991, A\&A 244, L5
\bibitem[1995]{lsl95}
        Lamers H.J.G.L.M., Snow T.P., Lindholm D.M., 1995, ApJ 455, 269
\bibitem[1998]{la98}
        Lamers H.J.G.L.M., Zickgraf F.-J., de Winter D., Houziaux L., Zorec J.,
        1998, A\&A 340, 117
\bibitem[1982]{lu82}
        Lucy L.B., 1982, ApJ 255, 286
\bibitem[1999]{mae}
        Maeder A., 1999, A\&A 347, 185
\bibitem[1988]{ow88}
        Owocki S.P., Castor J.I., Rybicki G.B., 1988, ApJ 335, 914
\bibitem[1994]{ocb94}
        Owocki S.P., Cranmer R.C., 1994, ApJ 424, 887
\bibitem[1998]{ocg98}
        Owocki S.P., Cranmer R.C., Gayley K.G., 1998, Ap\&SS 260, 149  
\bibitem[1986]{ppk86}
        Pauldrach A., Puls J., Kudritzki R.P., 1986, A\&A 164, 86
\bibitem{}
        Puls J., Petrenz P., Owocki S.P., 1999 
        in {\it Variable and non-spherical stellar winds in
        luminous hot stars}, Lecture Notes in Physics, eds. B. Wolf et al.
        Springer: Heidelberg, p 131.
\bibitem[1999]{vink99}
        Vink J.S., de Koter A., Lamers H.J.G.L.M., 1999, A\&A  350, 181
\bibitem[1924]{zeipel}
        von Zeipel H., 1924, MNRAS 84, 665      
\bibitem[1992]{zick92}
        Zickgraf F.-J., 1992, in {\it Astronomical CCD Observing and
        Reduction Techniques},
        ed. Howell S.B., ASP Conference Series Vol. 22, 75
\bibitem[1989]{zsl89}
        Zickgraf F.-J., Schulte-Ladbeck R.E., 1989, A\&A 214, 274
\bibitem[1985]{zick85}
        Zickgraf F.-J., Wolf B., Stahl O., Leitherer C., Klare G., 1985, A\&A 143, 421



\end{thebibliography}
\end{document}